\documentstyle[12pt]{article}

\begin{document}
\thispagestyle{empty}
{\baselineskip=12pt
\hfill HUTP-96/A006 

\hfill hep-th/9603059 

\vspace{1.0cm}}
\centerline{\large \bf Duality and $4d$ 
String Dynamics\footnote{Research supported in part by the 
Harvard Society of Fellows.}}
\bigskip
\bigskip
\centerline{\large Shamit Kachru \footnote{Email: kachru@string.harvard.edu}}
\medskip
\centerline{\it Harvard University, Cambridge, MA 02138, USA}
\bigskip
\bigskip
\bigskip
\centerline{{\it Presented at the Workshop ``Frontiers in Quantum 
Field Theory''}}
\centerline{{\it in Honor of the 60th birthday of Keiji Kikkawa}}
\centerline{{\it Osaka, Japan \quad December 1995}}
\vskip 1.0 truein
\parindent=1 cm

\begin{abstract}

We review some examples of heterotic/type II string duality which
shed light on the infrared dynamics of string compactifications with
N=2 and N=1 supersymmetry in four dimensions.

\end{abstract}
\vfil\eject
\setcounter{page}{1}
\section{Introduction}
\vglue0.4cm

I am very happy to have the opportunity to speak about strong/weak 
coupling duality on
this occasion honoring the 60th birthday of Professor Keiji Kikkawa.  
His own foundational work on T-duality \cite{KY}, the 
worldsheet analogue of S-duality,  
was in many ways instrumental in inspiring the recent 
developments in nonperturbative string theory.

Strong-weak coupling dualities now allow us to determine the strong coupling
dynamics of string vacua with $N \geq 4$ supersymmetry in four dimensions
\cite{Schwarz}.
It is natural to ask if this progress in our understanding of string theory
can be extended to the more physical vacua with less supersymmetry.
For N=2 theories in four dimensions, 
quantum corrections significantly modify the mathematical structure of
the moduli space of vacua, as well as the
physical interpretation of its apparent singularities. 
This was beautifully demonstrated in the field theory case in 
\cite{SeiWit} and it has more recently become possible to compute the
exact quantum moduli spaces for N=2 string compactifications as 
well \cite{KV,FHSV}. 
This constitutes the subject of the first part of my talk.

Of course, the case of most physical interest is $N \leq 1$ theories.
In the second part of my talk, I discuss examples of dual heterotic/type II
string pairs
where the heterotic theory is expected to exhibit nonperturbative dynamics
which may fix the dilaton and break supersymmetry \cite{KS}.  The type II 
dual manages to reproduce the qualitative features expected of the
heterotic side at tree level.  It is to be hoped that further work
along similar lines will result in a better understanding of
supersymmetry breaking in string theory.

The first part of this talk is based on joint work with C. Vafa, and the
second part of this talk is based on joint work with E. Silverstein.

\section{N=2 Gauge Theory and String Compactifications}

Recall that the N=2 gauge theory with gauge group $SU(2)$ is the theory of
a single N=2 vector multiplet consisting of a vector $A^{\mu}$, 
two Weyl fermions $\lambda$ and $\psi$,
and a complex scalar field $\phi$, all in the adjoint representation of 
$SU(2)$. 
In N=1 language, this is a theory of an N=1 vector multiplet
$(\lambda, A^{\mu})$ coupled to an N=1 chiral multiplet $(\phi, \psi)$.
The scalar potential of the theory is determined by supersymmetry to
be                
\begin{equation}
V(\phi) = {1\over g^{2}} [\phi, \phi^{+}]^{2}
\end{equation}
We see that $V$ vanishes as long as we take $\phi = diag (a,-a)$, so there
is a moduli space of classical vacua parameterized by the
gauge invariant parameter $u = tr(\phi^{2})$.

At generic points in this moduli space ${\cal M}_{v}$ of vacua, there is
a massless N=2 U(1) vector multiplet $A$.  The leading terms in its effective
lagrangian are completely determined in terms of a single holomorphic
function $F(A)$, the prepotential:
\begin{equation}
L \sim \int d^{4}\theta {\partial F \over {\partial A}}\bar A +
\int d^{2}\theta {\partial^{2} F \over{\partial A^{2}}}W_{\alpha}W^{\alpha}
 + c.c.
\end{equation} 
The first term determines, in N=1 language, the Kahler potential (and hence
the metric on ${\cal M}_{v}$) while the second term determines the
gauge coupling as a function of moduli.

In \cite{SeiWit} the exact form of $F$ including 
instanton corrections was determined.  
In addition, the masses of all of the BPS saturated particles were 
computed.  This was reviewed in great detail in several other talks
at this conference, so I will not repeat the solution here.  It will
suffice to say that  
the crucial insight is that the
singular   
point $u=tr(\phi^{2}) = 0$ where $SU(2)$ gauge symmetry is restored in
the classical theory splits, in the quantum theory, into two 
singular points
$u = \pm \Lambda^{2}$, where a monopole and a dyon become massless. 

In this talk our interest 
is not really in $N=2$ gauge theories but in the string
theories which reduce to $N=2$ gauge theories in the infrared. 
There are two particularly simple classes of $d=4, N=2$ supersymmetric
string compactifications.  One obtains such theories from Type II (A or B)
strings on Calabi-Yau manifolds, and from heterotic strings on $K_{3}
\times T^{2}$ (with appropriate choices of instantons on the $K_3$).  
Here we briefly summarize some basic properties of these theories. 

Type IIA strings on a Calabi-Yau threefold $M$ give rise to a four-dimensional
effective theory with $n_v$ vector multiplets and $n_{h}$ hypermultiplets
where 
\begin{equation}
n_{v} = h^{1,1}(M), ~~n_{h} = h^{2,1}(M) + 1 
\end{equation}
The $+1$ in $n_{h}$ corresponds to the fact that for such type II
string compactifications, the $\it dilaton$ is in a hypermultiplet.

The vector fields in such a theory are Ramond-Ramond U(1)s, so there
are no charged states in the perturbative string spectrum.  Furthermore, 
because of the theorem of de Wit, Lauwers, and Van Proeyen \cite{dLVP}
which forbids couplings of vector multiplets to neutral hypermultiplets
in N=2 effective lagrangians, the dilaton does not couple to the vector  
moduli.  This means that there are no perturbative or nonperturbative
corrections to the moduli space of vector multiplets.
On the other hand the moduli spaces of hypermultiplets are expected to
receive highly nontrivial corrections, including ``stringy'' corrections
with $e^{-1/g}$ strength \cite{BBS}.
 
One interesting feature of the moduli spaces of vector multiplets in
such theories is the existence of conifold points at finite distance
in the moduli space.  At such points the low energy effective theory
becomes singular (e.g., the prepotential develops a logarithmic
singularity) \cite{CDGP}.  This phenomenon is reminiscent of 
the singularities in the prepotential which occur at the ``massless
monopole'' points in the Seiberg-Witten solution of N=2 gauge theory,
singularities which are only present because one has integrated out a
charged field which is becoming massless.  In the case at hand, in fact,
one can show that there are BPS saturated states (obtained by wrapping
2-branes around collapsing 2-cycles) which become massless and which
are charged under (some of) the Ramond-Ramond $U(1)$s \cite{Strominger}.  These
explain the singularity in the prepotential.   In fact at special
such points, where enough charged fields (charged under few enough
$U(1)$s) become massless, one can give them VEVs consistent with
D and F flatness.  This results in new ``Higgs branches'' of the moduli
space. These new branches correspond to string compactifications on
different Calabi-Yau manifolds, topologically distinct from $M$
\cite{GMS}, and there is evidence that all Calabi-Yau 
compactifications may be connected
in this manner \cite{Cornell,Texas}. 

The other simple way of obtaining an N=2 theory in four dimensions from
string theory is to compactify the heterotic string (say $E_8\times E_8$)
on $K_3 \times T^2$.  Because of the Bianchi identity
\begin{equation}
dH = Tr(R\wedge R) - Tr(F\wedge F)
\end{equation}   
one must embed 24 instantons in the $E_8 \times E_8$ in order
to obtain a consistent theory.
An $SU(N)$ k-instanton on $K_3$ comes with $Nk + 1 - N^2$ hypermultiplet moduli
(where $k\geq 2N$), and $K_3$ comes with 20 hypermultiplet moduli
which determine its size and shape.  
Embedding an $SU(N)$ instanton in $E_8$ breaks the observable low
energy gauge group to the maximal subgroup of $E_8$ which commutes with
$SU(N)$ ($E_7$ for N=2, $E_6$ for N=3, and so forth). 

In addition, there are three $U(1)$ vector multiplets associated with 
the $T^2$.  Their scalar components are the dilaton $S$ and the
complex and kahler moduli $\tau$ 
and $\rho$ of the torus (both of which live on the upper half-plane
$H$ mod $SL(2,Z)$).  At special points in the moduli space
the $U(1)^2$ associated with $\tau$ and $\rho$ is enhanced to
a nonabelian gauge group:
\begin{equation}
\tau = \rho \rightarrow SU(2)\times U(1),~~ \tau=\rho=i \rightarrow SU(2)^2, 
~~\tau = \rho = 1/2 + i{\sqrt 3}/2 \rightarrow SU(3)
\end{equation}

Because the dilaton lives in a vector multiplet in such compactifications,
the moduli space of vectors is modified by quantum effects.  On the other
hand, the moduli space of hypermultiplets receives neither perturbative nor
nonperturbative corrections. 

An interesting feature of the heterotic ${\cal M}_{v}$ 
is the existence of special points where the
classical theory exhibits an enhanced gauge symmetry (as described
above for the compactification on $T^2$).
Sometimes by    
appropriate passage to a Higgs or Coulomb phase, such enhanced gauge
symmetry points link moduli spaces of N=2 heterotic theories which
have different generic spectra (for some examples see
\cite{KV,AFIQ}).  It is natural to conjecture that              
such transitions connect all heterotic N=2 models, in much the same way
that conifold transitions connect Calabi-Yau compactifications of type II
strings.

\section{N=2 String-String Duality}

>From the brief description of heterotic and type II N=2 vacua in the
previous section, it is clear that a duality relating the two classes of
theories would be extremely powerful.  If one were to find a model with dual
descriptions as a compactification of the Type IIA string on $M$ and 
the heterotic string on $K_3 \times T^{2}$, one could compute the exact
prepotential for ${\cal M}_{v}$ from the Type IIA side (summing up what from
the heterotic perspective would be an infinite series of instanton 
corrections).  Similarly, one would get exact results for ${\cal M}_{h}$
from the heterotic side --  this would effectively compute the $e^{-1/g}$
corrections expected from the IIA perspective.  In fact, such a duality 
has been found to occur in several examples in \cite{KV,FHSV}.

One of the simplest examples is as follows.
Consider the heterotic string compactified to eight dimensions on $T^{2}$ 
with $\tau = \rho$.  Further compactify on a $K_{3}$, satisfying the
Bianchi identity for the $H$ field by embedding 
$c_{2}=10$ $SU(2)$ instantons in each $E_8$ and a $c_{2}=4$ $SU(2)$ instanton
into the ``enhanced'' $SU(2)$ arising from the $\tau=\rho$ torus.
After Higgsing the remaining $E_7$ gauge groups one is left with a generic
spectrum of 129 hypermultiplets and 2 vector multiplets.
The 2 vectors are $\tau$ and the dilaton $S$ -- when $\tau = i$, one expects
an $SU(2)$ gauge symmetry to appear (the other $SU(2)$ factor that 
would normally
appear there has been broken in the compactification process).

This tells us that if there is a type IIA dual compactification on a 
Calabi-Yau $M$, then the Betti numbers of $M$ must be 
\begin{equation} 
h_{11}(M) = 2, ~~h_{21}(M) = 128
\end{equation} 
There is a known candidate manifold with these Betti numbers -- the
degree 12 hypersurface in $WP^{4}_{1,1,2,2,6}$ defined by the vanishing of
$p$
\begin{equation}
p = z_{1}^{12} + z_{2}^{12} + z_{3}^{6} + z_{4}^{6} + z_{5}^{2} + ....
\end{equation}
This manifold has in fact been studied intensively as a simple example of
mirror symmetry in \cite{Hosono,Morrison}.   

The mirror manifold $W$ has $h_{11}(W) = 128, h_{21}(W) = 2$.  The conjecture
that IIA on $M$ is equivalent to the heterotic string described above 
implies that IIB on $W$ is also equivalent to that heterotic
string.  The structure of the moduli space of vector 
multiplets of the heterotic string should be
$\it exactly$ given by the classical (in both sigma model and string
perturbation theory) moduli space of complex structures of $W$. 

The mirror manifold can be obtained by orbifolding
$p=0$ by the maximal group of phase symmetries which preserves the
holomorphic three-form \cite{GP}.  Then the two vector moduli are represented
by $\psi$ and $\phi$ in the polynomial 
\begin{equation}
p = z_{1}^{12} + z_{2}^{12} + z_{3}^{6} + z_{4}^{6} + z_{5}^{2} - 12 \psi
z_{1}z_{2}z_{3}z_{4}z_{5} - 2\phi z_{1}^{6} z_{2}^{6} 
\end{equation} 
It is also useful, following \cite{Hosono}, to introduce
\begin{equation}
x = {-1\over 864} {\phi\over \psi^{6}}, ~~y = {1\over \phi^{2}}
\end{equation} 
These are the convenient ``large complex structure'' coordinates on the
moduli space of vector multiplets for the IIB string.                 

In order to test our duality conjecture, we should start by checking
that the IIB string reproduces some qualitative features that we expect
of the heterotic ${\cal M}_{v}$.  For example, $\tau = i$ for weak coupling
$S\rightarrow \infty$ is an $SU(2)$ point.  There should therefore be a  
singularity of ${\cal M}_{v}$ at this point which splits, as one turns on 
the string coupling, to $\it two$ singular points (where monopoles/dyons
become massless), as in the case of pure $SU(2)$ gauge theory.

The ``discriminant locus'' where the IIB model becomes singular is given
by
\begin{equation}
(1-x)^{2} - x^{2} y = 0
\end{equation}
So we see that as a function of $y$ for $y \neq 0$ there are two solutions
for $x$ and as $y \rightarrow 0$ they merge to a single singular point
$x=1$.  This encourages us to identify $x=1, y=0$ with $\tau =i, S \rightarrow
\infty$ of the heterotic string -- the $SU(2)$ point.  The metric on the
moduli space for $y$ at $y=0$ and $S$ at weak coupling also agree if one 
makes the identification $y\sim e^{-S}$.

There is also a remarkable observation in 
\cite{Morrison} that the mirror map, restricted to $y=0$, is given by
\begin{equation}
x = {j(i)\over j(\tau_{1})}
\end{equation}
where $\tau_{1}$ is one of the coordinates on the Kahler cone of $M$.
Here $j$ is the elliptic j-function mapping $C$ onto $H/SL(2,Z)$.
This tells us that the classical heterotic $\tau$ moduli space,
which is precisely $H/SL(2,Z)$, is embedded in the moduli space of $M$
at weak coupling precisely as expected from duality.
In fact using the uniqueness of special coordinates up to rotations,
one can find the exact formula expressing the IIB coordinates
$(x,y)$ in terms of the heterotic coordinates $(\tau,S)$. 

Of course with this map in hand there are now several additional things
one can check.  The tests which have been performed in 
\cite{KV,KLT,AGNT,KKLMV} include
\medskip

\noindent 1) A matching 
of the expected loop corrections to the heterotic prepotential
with the form of the tree-level exact Calabi-Yau prepotential.
\medskip

\noindent 2) A test that 
the g-loop F-terms computed by the topological partition
functions $F_g$  on the type II side (which include e.g. $R^{2}$ and other
higher derivative terms) are reproduced by appropriate (one-loop!)
computations on the heterotic side.
\medskip

\noindent 3) A demonstration that in an appropriate double-scaling limit,
approaching the $\tau = i$, $S \rightarrow \infty$ point of the heterotic
string while taking $\alpha^{'} \rightarrow 0$, the IIB prepotential
reproduces the exact prepotential of $SU(2)$ gauge theory 
(including Yang-Mills instanton effects) computed in \cite{SeiWit}.

\medskip
These tests give very strong evidence in favor of the conjectured duality.
Given its veracity, what new physics does the duality bring into reach?
\medskip

\noindent $\bullet$ One now has examples of four-dimensional theories
with exactly computable quantum gravity corrections. In the example
discussed above, the
Seiberg-Witten prepotential which one finds in an expansion
about $\tau = i, S \rightarrow \infty$
receives gravitational corrections which
are precisely computable as a power series in $\alpha^{'}$. 
\medskip

\noindent $\bullet$ On a more conceptual level, the approximate duality of
\cite{SeiWit} between a microscopic $SU(2)$ theory (at certain points in its
moduli space) and a $U(1)$ monopole/dyon theory is promoted to an
$\it exact$ duality, valid at all wavelengths, between heterotic and
type II strings. 
\medskip

\noindent $\bullet$ There is evidence that at strong heterotic coupling,
new gauge bosons and charged matter fields appear, sometimes giving rise
to new branches of the moduli space \cite{KMP,KM}.
\medskip

\noindent$\bullet$ The $e^{-1/g}$ corrections to the hypermultiplet moduli
space of type II strings are in
principle exactly computable using duality (and may be of some mathematical
interest).

\medskip
One might wonder what is special about the Calabi-Yau manifolds which are
dual to weakly coupled heterotic strings.
In fact it was soon realized that the examples of duality in \cite{KV} involve
Calabi-Yau manifolds which are $K_3$ fibrations \cite{KLM}.  That is, locally 
the Calabi-Yau looks like $CP^{1}\times K_{3}$.
In fact, one can prove that if the type IIA string on a Calabi-Yau $M$
(at large radius) 
is dual to a weakly coupled heterotic string, then $M$ must be a $K_3$
fibration \cite{AL}.

To make this more concrete,
in the example of the previous section, we saw $M$ was defined by the
vanishing of
\begin{equation}
p = z_{1}^{12} + z_{2}^{12} + z_{3}^{6} + z_{4}^{6}  + z_{5}^{2} + ...
\end{equation}
in $WP^{4}_{1,1,2,2,6}$.  Set 
$z_{1}=\lambda z_{2}$ and define $y=z_{1}^{2}$ (which is an allowed
change of variables since an identification on the $WP^{4}$ takes 
$z_{1} \rightarrow -z_{1}$ without acting on $z_{3,4,5}$).  Then the
polynomial becomes (after suitably rescaling to absorb $\lambda$)
\begin{equation}
p = y^{6} + z_{3}^{6} + z_{4}^{6} + z_{5}^{2} + ... 
\end{equation}
which defines a $K_{3}$ surfaces in $WCP^{3}_{1,1,1,3}$.  The choice
of $\lambda$ in $z_{1}=\lambda z_{2}$ is a point on $CP^{1}$, and the
$K_{3}$ for fixed choice of $\lambda$ is the fiber. 

It is not surprising that $K_3$ fibrations play a special role in 
4d N=2 heterotic/type II duality.  Indeed the most famous example of
heterotic/type II duality is the 6d duality between heterotic strings on 
$T^{4}$ and type IIA strings on $K_{3}$ \cite{HT,Witten}.  If 
one compactifies the type IIA
string on a CY threefold which is a $K_3$ fibration, and simultaneously
compactifies the heterotic string on a $K_{3}\times T^{2}$ where the $K_3$
is an elliptic fibration, then locally one can imagine taking the bases
of both fibrations to be large and obtaining in six dimensions an example
of the well-understood 6d string-string duality \cite{VW}.  This picture is not
quite precise because of the singularities in the $K_3$ fibration,
but it does provide an intuitive understanding of the special role of
$K_3$ fibrations.

\section{N=1 Duality and Gaugino Condensation}
 
Starting with an $N=2$ dual pair of the sort discussed above, one can try
to obtain an $N=1$ dual pair by orbifolding both sides by freely acting
symmetries.   This strategy was used in \cite{VW,HLS} where several
examples with trivial infrared dynamics were obtained.  Here we will
find that examples with highly nontrivial infrared dynamics can also
be constructed \cite{KS}.

Our starting point is 
an N=2 dual pair 
(IIA on a Calabi-Yau $M$ and heterotic on $K_{3}\times T^{2}$)
where the heterotic gauge group takes the form
\begin{equation}
G ~=~E_{8}^{H} \otimes E_{7}^{obs}\otimes ...
\end{equation}
$H$ denotes the hidden sector and $obs$ the observable sector.
We will first discuss the technical details of the $Z_2$ symmetry by 
which we can orbifold both sides
to obtain an $N=1$ dual pair, and then we discuss the physics of the
duality.

Orbifold the heterotic side by the Enriques involution 
acting on $K_3$ 
and a total reflection on the $T^{2}$.  This acts on the base of the elliptic
fibration $(z_{1},z_{2})$ by
\begin{equation}
(z_{1}, z_{2}) ~ \rightarrow ~ (\bar z_{2}, - \bar z_{1})
\end{equation}
taking $CP^{1} \rightarrow RP^{2}$.
In addition, we need to choose a lifting of the orbifold group to the gauge
degrees of freedom.

We do this as follows:

\noindent $\bullet$ Put a modular invariant embedding into the ``observable''
part of the gauge group alone.

\noindent $\bullet$ Embed the translations which generate the $T^2$ into
$E_{8}^{H}$, constrained by maintaining level-matching and the relations of
the space group.  For example one could take Wilson lines $A_{1,2}$ 
along the $a$
and $b$ cycle of the $T^{2}$ given by
\begin{equation}
A_{1} = {1\over 2} (0,0,0,0,1,1,1,1),~~~A_{2} = {1\over 2}(-2,0,0,0,0,0,0,0)
\end{equation}
Here $A_{1,2} = {1\over 2}L_{1,2}$ where $L_{1,2}$ are vectors in the 
$E_8$ root lattice.  These Wilson lines break the $E_{8}^{H}$ gauge
symmetry to $SO(8)_{1} \otimes SO(8)_{2}$. 

\medskip
How does the $Z_2$ map over to the type II side?
>From the action
\begin{equation}
(z_{1},z_{2}) ~\rightarrow ~(\bar z_{2}, -\bar z_{1})
\end{equation}   
on the $CP^{1}$ base (which is common to both the heterotic and type II sides),
we infer that the $Z_2$ must be an antiholomorphic, orientation-reversing 
symmetry of the Calabi-Yau manifold $M$.
To make this a symmetry of the type IIA string theory, we must simultaneously
flip the worldsheet orientation, giving us an ``orientifold.''  
In such a string theory, one only includes maps $\Phi$ of the worldsheet
$\Sigma$ to spacetime $M/Z_{2}$ if they satisfy
\begin{equation}
\Phi^{*}(w_{1}(M/Z_{2})) = w_{1}(\Sigma)
\end{equation} 
where $w_1$ is the first Stieffel-Whitney class.

We know from 6d string-string duality that the Narain lattice $\Gamma^{20,4}$
of heterotic string compactification on $T^4$ maps to the integral cohomology
lattice of the dual $K_{3}$.  This means that we can infer from the action
of the $Z_2$ on the heterotic gauge degrees of freedom, what the action of
the $Z_2$ must be on the integral cohomology of the $K_3$ fiber on the IIA
side.
Since we are frozen on the heterotic side at a point with $SO(8)^{2}$ gauge
symmetry in the hidden sector, the dual $K_3$ must be frozen at its singular
enhanced gauge symmetry locus.  

The $K_3$ dual to heterotic enhanced gauge symmetry $G$ has rational
curves $C_i$, $i = 1,...,rank(G)$ shrinking to zero area (with the 
associated $\theta_{i}=0$ too).  It is easy to see, e.g. from Witten's
gauged linear sigma model that in this situation the type II theory
indeed exhibits an extra $Z_2$ symmetry.  The bosonic potential of
the relevant gauged linear sigma model (for the case of a single
shrinking curve) is given by 
$$V=
{1\over{2e^2}}
\sum_i\biggl\{\biggl(\bigl[\sum_\alpha Q_i^\alpha(|\phi^i_\alpha|^2
-|\tilde\phi^i_\alpha|^2)\bigr]-r_i^0\biggr)^2$$
$$+\biggl(Re(\sum_\alpha\phi^i_\alpha\tilde\phi^i_\alpha)-r_i^1\biggr)^2
+\biggl(Im(\sum_\alpha\phi^i_\alpha\tilde\phi^i_\alpha)-r_i^2\biggr)^2
\biggr\}$$ 
$$+{1\over 2}\sum_i\bigl[\sum_\alpha Q^{\alpha~2}_i
(|\phi^i_\alpha|^2+|\tilde\phi^i_\alpha|^2)\bigr]|\sigma_i|^2$$ 
Here the $\phi$s represent the $K_3$ coordinates while $r$ parametrizes
the size of the curve and $\sigma$ is the Kahler modulus.
Precisely when $\vec r \rightarrow 0$, the model has the $Z_2$ symmetry
$\phi \rightarrow -\tilde \phi$, $\sigma \rightarrow - \sigma$.  
Orbifolding
by this $Z_2$ then freezes the $K_3$ at its enhanced gauge symmetry locus,
as expected.

What is the physics of the dual pairs that one constructs in this manner?
In the heterotic string, when there is a hidden sector pure gauge group
\begin{equation}
G^{hidden} = \Pi ~G^{b}
\end{equation}
one expects gaugino condensation to occur.  This induces an effective
superpotential
\begin{equation} 
W = \sum ~h_{b}~ \Lambda_{b}^{3}(S)
\end{equation}
where $\Lambda_{b}(S) \sim e^{-\alpha_{b} S}$ and $\alpha_{b}$ is 
related to the
beta function for the running $G_b$ coupling.  It was realized early on
\cite{Krasnikov,DKLP} that in such models (with more than one hidden factor)
one might expect both stabilization of the dilaton and supersymmetry 
breaking.
It has remained a formidable problem to determine which (if any) such models
actually do have a stable minimum at weak coupling with broken supersymmetry.

For now, we will be content to simply understand how the $\it qualitative$
structure of the heterotic theory (e.g. the gaugino-condensation induced
effective superpotential) is reproduced by the type II side.
This is mysterious because the type II N=2 theory we orientifolded had only
abelian gauge fields in its spectrum, so we need to reproduce the strongly
coupled nonabelian dynamics of the heterotic string with an $\it abelian$
gauge theory on the type II side.

The heterotic orbifold indicates the spectrum of the string theory as
$g_{het} \rightarrow 0$.  The
heterotic dilaton $S$ maps to the radius $R$ of the $RP^{2}$ base of the
type II orientifold
(recall one obtains the $RP^2$ by orbifolding the
base $P^1$ of the $K_3$ fibration)
\begin{equation}
S_{het} \leftrightarrow  R_{RP^{2}}
\end{equation} 
The purported stable vacuum of the heterotic theory should then be expected
to lie at large radius for the base, and on the (orientifold of the) conifold
locus dual to enhanced gauge symmetry.  There are two crucial features of this
locus:
\medskip

\noindent 1) The $RP^2$ base has $\pi_{1}(RP^{2}) = Z_{2}$.  So a state
projected out in orientifolding the N=2 theory will have a massive version
invariant under the $Z_2$.  Say $\beta \in \pi_{1}(RP^{2})$ 
is the nontrivial element.  Take $x$ a coordinate along an appropriate
representative of $\beta$ -- a representative can be obtained by taking 
the image of a great circle on the original base $P^1$ after 
orientifolding. 
Then if the original non-invariant vertex operator was $V$, a
new invariant vertex operator is given adiabatically by
\begin{equation}
V^{\prime} = e^{ix \over R} V 
\end{equation}
The $Z_2$ takes $x$ to $x + \pi R$ and therefore $V^{\prime}$ is invariant
if $V$ was not.  In particular this gives us massive versions of the
scalars $a^{i}_{b,D}$ in the N=2 vector multiplets for $G^{b}$ with masses
\begin{equation}
M_{a} \sim {1\over R^{2}}
\end{equation}
Effectively, for very large $R$, one is restoring the original N=2 
supersymmetry. 
\medskip 

\noindent 2) The low energy theory for IIA at the conifold locus contains
massless $\it solitonic$ states \cite{Strominger}.  One can see that
they survive the N=2 $\rightarrow$ N=1 orientifolding by examining 
the behavior of the gauge couplings \cite{VW}.  These extra solitonic
states play the role of the ``monopole hypermultiplets'' $M_{i}^{b}, 
\tilde M_{i}^{b}$ of the N=2 theory. 

\medskip
These two facts taken together imply that as $R \rightarrow \infty$ there
is an effective superpotential
\begin{equation}
W_{II} = \sum_{b} \left( m_{b}u^{b}_{2}(a^{i}_{b,D},R) +
\sum_{i=1}^{rank(b)} M_{i}^{b} a^{i}_{b,D}\tilde M^{b}_{i} \right)
\end{equation} 
where $u^{b}_{2}$ is the precise analogue of $u$ of \S2 for $G^b$ and 
its functional dependence on $R$ can be found from the $N=2$ dual pair. 
As we'll now discuss, this structure
\medskip 

\noindent a) Allows us to reproduce the gaugino-condensation induced
effective superpotential of the heterotic side.  
\medskip

\noindent b) Implies $\langle M \rangle \neq 0$, suggesting a geometrical
description of the type II side by analogy with N=2 conifold transitions.

\medskip
To see a), recall how the physics of N=1 $SU(2)$ gauge theory is
recovered from the N=2 theory in \cite{SeiWit}.  One can obtain the
N=1 theory by giving a bare mass to the adjoint scalar in the N=2
vector multiplet and integrating it out.  In the vicinity of the
monopole points this means there is an effective superpotential 
\begin{equation}
W = m u(a_{D}) + \sqrt{2} a_{D} M \tilde M
\end{equation}
Using the equations of motion and D-flatness, one finds
\begin{equation}
\vert \langle M \rangle \vert = 
\vert \langle \tilde M \rangle \vert 
= ( -mu^{\prime}(0)/\sqrt{2})^{1/2},~~a_{D} = 0 
\end{equation}
The monopoles condense and given a mass to the (dual) $U(1)$ gauge field
by the Higgs mechanism, leaving a mass gap.  Two vacua arise in this way --
one at each of the monopole/dyon points -- in agreement with the 
Witten index computation for pure $SU(2)$ gauge theory.  
 
In our case, we expect that condensation of the massless solitons will lead
to the gaugino condensation induced superpotential (and, perhaps, supersymmetry
breaking).  To see this we must expand $W_{II}$ in $a^{i}_{b,D}$ 
in anticipation of finding
a minimum at small $a_{D}$ (since the minimum was at $a_{D} = 0$ in the global
case). 
Using
\begin{equation}
u_{2}(a^{i}_{b,D},R)) = e^{i\gamma^{b}}\lambda_{b}^{2}(R) + \cdots  
\end{equation}
(where $\gamma^{b}$ comes from the phase of the gaugino condensate)
as well as the matching condition
\begin{equation}
m_{b}\Lambda_{b,high}^{2} = \Lambda_{b,low}^{3}
\end{equation}
one obtains from integrating out the massive adjoint scalar, one sees 
\begin{equation}
W_{II} = \sum_{b} e^{i\gamma^{b}}\Lambda_{b}^{3}(S) + \cdots
\end{equation} 
Simply minimizing the supergravity scalar potential
\begin{equation} 
V = e^{K}(D_{i}W G^{i\bar j}D_{\bar j}W - 3 \vert W \vert^{2}) + 
{1\over 2}g^{2}D^{2}
\end{equation}
we also find that 
\begin{equation}
\langle M_{i}^{b} \tilde M_{i}^{b} \rangle = -h_{b}m_{b}u_{2,i}^{b}(S) - K_{i}
W
\end{equation}
That is, the ``wrapped two-branes'' which give us the massless monopoles have
condensed, in accord with the global result.
So integrating out the massive $M, \tilde M$ and adjoint scalar degrees of
freedom yields the same form of bosonic potential that we expect from
gaugino condensation on the heterotic side.

In summary, we have argued that the type II dual description of the effects
of gaugino condensation involves a mass perturbation breaking N=2 
supersymmetry.  One cannot add mass terms by hand in string theory: The
type II orientifold produces the requisite massive mode as a Kaluza-Klein
excitation of the original N=2 degrees of freedom that were projected out.

One intriguing feature of the IIA vacuum is the nonzero VEVs for the wrapped
two-branes $M, \tilde M$.  In the N=2 context $\langle M \rangle \neq 0 
\rightarrow$ conifold transition.  There is a well known geometrical 
description
of the conifold points.  For example, in the IIB theory the conifold 
in vector multiplet moduli space is obtained by going to a point in 
${\cal M}_{complex}$ where there is a cone over $S^{3}\times S^2$
in the Calabi-Yau.  One can either ``deform the complex structure'' (return
to the Coulomb phase, in physics language) by deforming the tip of the
cone into an $S^3$, or one can do a ``small resolution'' and blow the tip
of the cone into an $S^2$.  The latter corresponds to moving to a new
Higgs phase, in the N=2 examples \cite{GMS}.

It was noted long ago by Candelas, De La Ossa, Green, and Parkes \cite{CDGP}
that at a $\it generic$ conifold singularity such a small resolution does
not produce a Kahler manifold.  They speculated that such nonKahler resolutions
might correspond to supersymmetry breaking directions.  It is natural
to suggest that we might be seeing a realization of that idea by duality.
The analogy with N=2 conifold transitions suggests that
$\langle M \rangle \neq 0  \rightarrow$ nonKahler resolution.  One can hope
that this will provide a useful dual view of supersymmetry breaking in 
string theory.

\vfill\eject

\end{document}